\documentclass[aps,prl,preprint,superscriptaddress]{revtex4}
\usepackage{graphicx}

\usepackage{graphicx}
\usepackage{color}
\usepackage{amsfonts}
\begin{document}
\title{High-temperature behaviour of supported graphene: electron-phonon coupling and substrate-induced doping}

\author{S\o ren~Ulstrup}
\affiliation{Department of Physics and Astronomy, Interdisciplinary Nanoscience Center, Aarhus University,
8000 Aarhus C, Denmark}
\author{Marco~Bianchi}
\affiliation{Department of Physics and Astronomy, Interdisciplinary Nanoscience Center, Aarhus University,
8000 Aarhus C, Denmark}
\author{Richard~Hatch}
\affiliation{Department of Physics and Astronomy, Interdisciplinary Nanoscience Center, Aarhus University,
8000 Aarhus C, Denmark}
\author{Dandan~Guan}
\affiliation{Department of Physics and Astronomy, Interdisciplinary Nanoscience Center, Aarhus University,
8000 Aarhus C, Denmark}
\author{Alessandro~Baraldi}
\affiliation{Physics Department and CENMAT, University of Trieste, 34127 Trieste, Italy}
\affiliation{IOM-CNR Laboratorio TASC, Area Science Park, 34149 Trieste, Italy}
\author{Dario Alf\`e}
\affiliation{Department of Earth Sciences, Department of Physics and Astronomy, TYC@UCL, and London Centre for Nanotechnology, University College London, Gower Street, London WC1E 6BT, United Kingdom}
\author{Liv Hornek\ae r}
\affiliation{Department of Physics and Astronomy, Interdisciplinary Nanoscience Center, Aarhus University,
8000 Aarhus C, Denmark}
\author{Philip~Hofmann}
\affiliation{Department of Physics and Astronomy, Interdisciplinary Nanoscience Center, Aarhus University,
8000 Aarhus C, Denmark}
\email[]{philip@phys.au.dk}

\date{\today}

%


\maketitle

\textbf{
One of the salient features of graphene is the very high carrier mobility that implies tremendous potential for use in electronic devices \cite{Avouris:2007}. Unfortunately, transport measurements find the expected high mobility only in freely suspended graphene \cite{Bolotin:2008}. When supported on a surface, graphene shows a strongly reduced mobility, and an especially severe reduction for temperatures above 200~K \cite{Morozov:2008,Chen:2008}. A temperature-dependent mobility reduction could be explained by scattering of carriers with phonons, but this is expected to be weak for pristine, weakly-doped graphene \cite{Calandra:2007,Bolotin:2008b}. The mobility reduction has therefore been ascribed to the interaction with confined ripples or substrate phonons \cite{Morozov:2008,Fratini:2008,Chen:2008}. Here we study the temperature-dependent electronic structure of supported graphene by angle-resolved photoemission spectroscopy, a technique that can reveal the origin of the phenomena observed in transport measurements. We show that the electron-phonon coupling for weakly-doped, supported graphene on a metal surface is indeed extremely weak, reaching the lowest value ever reported for any material. However, the temperature-dependent dynamic interaction with the substrate leads to a complex and dramatic change in the carrier type and density that is relevant for transport. Using \emph{ab initio} molecular dynamics simulations, we show that these changes in the electronic structure are mainly caused by fluctuations in the graphene-substrate distance.
}

Graphene's remarkable transport properties have been one reason for the tremendous interest in this material  \cite{Novoselov:2004,Novoselov:2005} and have been widely studied \cite{Das-Sarma:2011}. Transport measurements give direct access to the quantities that are eventually important for applications, such as the temperature-dependent carrier density and mobility. In such experiments, graphene is typically placed on insulating SiO$_2$ so that the carrier density can be changed by electric field gating. Placing graphene on SiO$_2$, however, has been shown to severely reduce the carrier mobility, especially above 200~K, i.e.  for the temperature range relevant of applications \cite{Morozov:2008,Chen:2008}. This  can be improved by choosing a flat and non-polar insulator as a substrate, such as hexagonal boron nitride \cite{Dean:2010}, but the microscopic mechanism of the mobility reduction is not yet well understood.  Here we address this issue using a combination of angle-resolved photoemission spectroscopy (ARPES) and \textit{ab initio} molecular dynamics, techniques that can give detailed information on the system's spectral properties and are thus complementary to transport measurements. 

So far, all ARPES investigations of the electron-phonon coupling in graphene have been carried out at a constant, low temperature. The determination of  the electron-phonon mass enhancement parameter $\lambda$  then relies on the observed energy dependence of the electronic self-energy near the Fermi energy $E_F$. For this approach to be applicable, the sample temperature has to be much lower than the relevant temperature for phonon excitations. For the reported results this is fulfilled with respect to graphene's very high Debye temperature \cite{Pozzo:2011}, but it might not be fulfilled if the  Bloch-Gr\"uneisen temperature sets the relevant temperature scale \cite{Efetov:2010}. For strongly doped graphene ($n \approx 10^{13}$~cm$^{-2}$), the electron-phonon scattering was found to be of intermediate strength with  $\lambda \approx 0.2 - 0.3$ \cite{Bostwick:2007,Bostwick:2007b,Bianchi:2010}. For weakly doped graphene, $\lambda$ appears to be much smaller \cite{Forti:2011}.
Here we employ a different approach to studying the electron-phonon coupling directly, by measuring the temperature-dependent self-energy for graphene supported on a metal surface. This necessitates that ARPES experiments be carried out up to high temperatures but it does not require assumptions about the relevant temperature scale for phonon excitations (Debye vs. Bloch-Gr\"uneisen). In fact, the determination of the relevant temperature scale for phonon excitations is a by-product of the analysis. We find that $\lambda$ is extremely small such that no temperature-induced mobility reduction would be expected for this system. However,  we also find unexpected temperature-induced changes in the electronic structure near the Fermi energy that, in a transport measurement, would entirely dominate the electron-phonon coupling effect. 
 
\begin{figure}
\begin{center}
\includegraphics[width=0.9 \columnwidth]{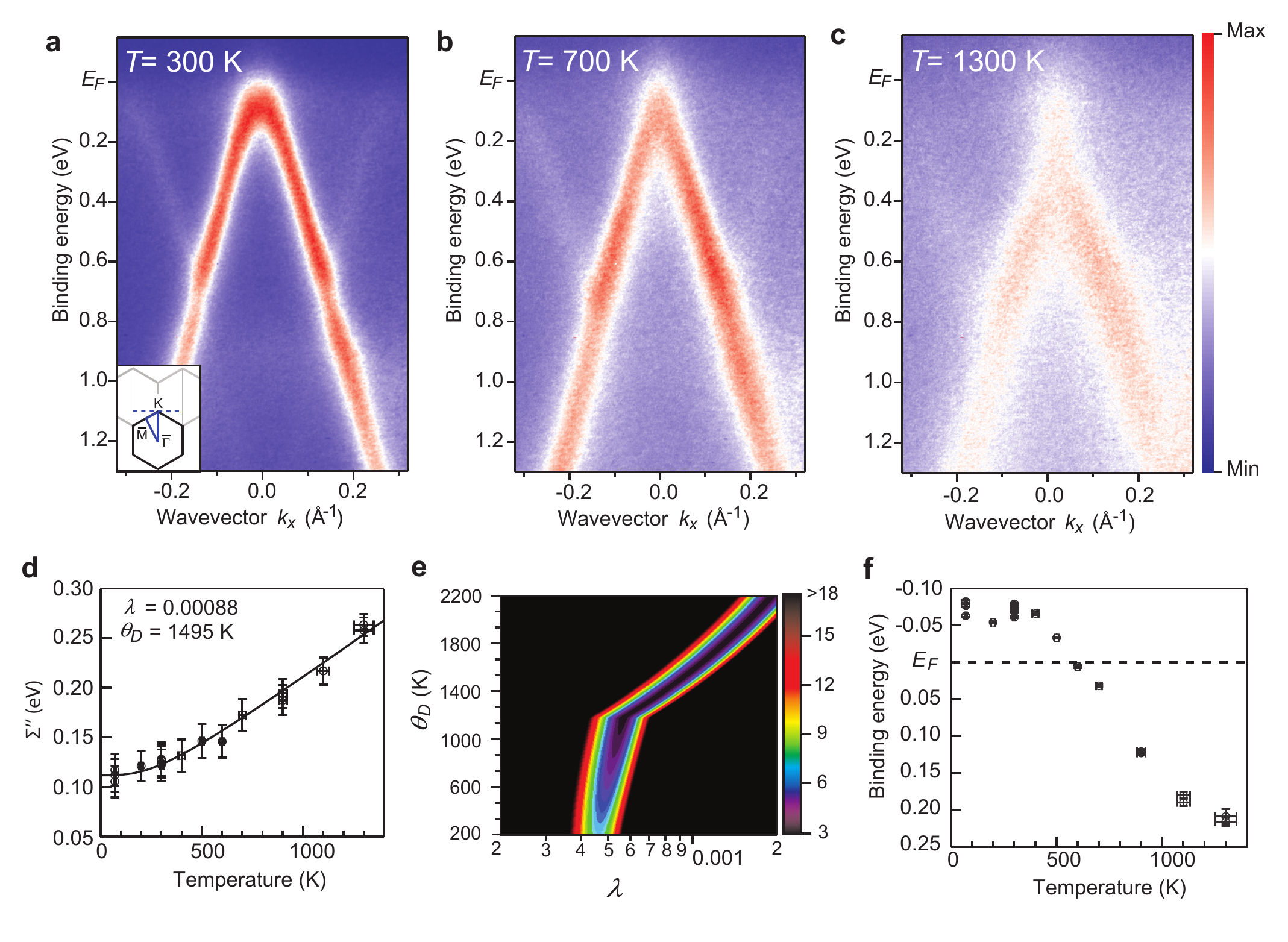}
\caption{Temperature-dependent electronic structure of graphene on Ir(111) determined by ARPES. (a)-(c) Spectra taken through the  $\bar{K}$ point of the Brillouin zone perpendicular to the $\bar{\Gamma}$-$\bar{K}$ direction (dashed line in the inset of (a)) for three different temperatures. The wavevector is measured relative to the Dirac point. (d) Imaginary part of the self-energy obtained as an average using the temperature-dependent momentum distribution linewidths between 250~meV and 550~meV below the Dirac point. The solid line is a result of fitting eq. (\ref{e1}) to the data using the given parameters. (e) $\chi^2$ value for the fit in (d) as a function of two fit parameters, the Debye temperature $\theta_D$ and the electron-phonon coupling strength $\lambda$. (f) Dirac point energy as a function of temperature, estimated from the extrapolated high binding energy dispersion.
\label{fig:1}}
\end{center}
\end{figure}

The temperature-dependent spectral function for graphene supported on Ir(111)  is shown in Figure \ref{fig:1}. The ARPES measurement of the electronic structure close to the Fermi energy $E_F$ and near the $\bar{K}$-point of the Brillouin zone is shown for three different temperatures in Figure \ref{fig:1}(a)-(c). The characteristic Dirac cone is easily identified, even for the highest temperature of 1300~K. In addition to the main Dirac cone, weak replicas and mini-gaps are evident. These are caused by the interaction with the substrate and the formation of a moir\'e superstructure \cite{Kralj:2011,Pletikosic:2009}. Remarkably, these features are clearly discernible even at the highest temperature. As the temperature is increased, several changes can be observed in the electronic structure. The first is the expected broadening of the features that is caused by the electron-phonon coupling. Given the very large temperature range of the measurements, this effect is relatively minor. The second and unexpected effect is a significant change of the doping. At 300~K the Dirac point of  graphene is located above the Fermi energy in agreement with earlier results  \cite{Kralj:2011,Pletikosic:2009}, but as the temperature increases, it shifts substantially and is clearly below the Fermi energy at 1300~K. Finally, the band structure at 1300~K does not show the expected Dirac cone-like dispersion but the spectral function around the Dirac point is broadened out and the situation resembles the observed onset of a gap-opening for disordered graphene \cite{Rotenberg:2007,Balog:2010}.

For a more detailed analysis of the electron-phonon coupling strength, we determine the linewidth of the momentum distribution curves (MDCs) averaged over binding energies  from 250~meV to 550~meV below the Dirac point as a function of temperature. From the average MDC linewidth and the (constant) group velocity $v$ of the band we infer the imaginary part of the self-energy $\Sigma''$ \cite{Hofmann:2009b} and plot this as a function of temperature in Figure \ref{fig:1}(d). In the high temperature limit, such data can directly yield the electron-phonon coupling strength $\lambda$ because $\Sigma''$ is a linear function of $T$, independent of the phonon spectrum \cite{McDougall:1995,Hofmann:2009}. For a metal, this high temperature limit is reached for $T$ higher than the Debye temperature $\Theta_D$. For graphene, this limit is not reached in our experiments, and it must also be kept in mind that the relevant temperature scale might not be set by $\Theta_D$ but rather by the  Bloch-Gr\"uneisen temperature $\Theta_{BG}$ that could be substantially lower \cite{Efetov:2010}. We thus have to employ the general expression   \cite{Grimvall:1981,Hofmann:2009}:
\begin{equation}
\Sigma''(T)=\pi\hbar\int_{0}^{\omega_{\mathrm{max}}}\alpha^{2}F(\omega^{\prime})[1-f(\omega-\omega^{\prime},T)+2n(\omega^{\prime},T)+f(\omega+\omega^{\prime},T)]d\omega^{\prime} +\Sigma''_0,
\label{e1}
\end{equation}
where $\hbar\omega$ is the hole energy, $\hbar\omega'$ is the phonon energy and
$f(\omega$,T) and $n(\omega,T)$ are the Fermi and Bose-Einstein distribution functions, respectively. $\Sigma''_0$ is a temperature-independent offset that accounts for electron-electron and electron-defect scattering. The integral extends over all phonon frequencies in the material. $\alpha^{2}F(\omega^{\prime})$ is the Eliashberg coupling function which we approximate by a 3D Debye model, i.e. 
\begin{equation}
\alpha^{2}F(\omega^{\prime}) =  \lambda ( \omega^{\prime}/ \omega_{D})^{2} = \lambda (\hbar \omega^{\prime}/ k_B \Theta_{D})^{2},
\label{e2}
\end{equation}
for ${\omega^{\prime} < \omega_{D}}$ and zero elsewhere.~\cite{Hellsing:2002}. A 3D model is chosen in view of the graphene-substrate interactions, but we note that choosing a 2D model does not significantly alter the results. 

In the further analysis, the data in Figure~\ref{fig:1}(d) are fitted using (\ref{e1}) and (\ref{e2}). This implies three fit parameters: $\Sigma''_0$, $\lambda$ and $\Theta_D$. We could choose to eliminate $\Theta_D$ from the fit by using an experimentally determined value (e.g. $\Theta_D = 1495$~K \cite{Pozzo:2011}). This, however, ignores the possibility that the actually relevant temperature scale is set by the Bloch-Gr\"uneisen temperature rather than the Debye temperature. We therefore choose to keep $\Theta_D$ in (\ref{e2}) as a free parameter and emphasise that the resulting $\Theta_D$ from the fit is then merely an effective measure of the temperature scale relevant for the electron-phonon scattering. It could be much lower than the actual Debye temperature determined from other experiments. In the fit, $\Theta_D$ and $\lambda$ are strongly correlated through (\ref{e2}) \cite{Kim:2005a}. Figure \ref{fig:1}(e) shows a plot of the resulting quality  of the fit ($\chi^2$) as a function of $\Theta_D$ and $\lambda$ and illustrates this correlation. We find equally good fits for a wide range of  $\Theta_D$ and $\lambda$ along the minimum of the contour, but only for values of  $\Theta_D\gtrsim1050$~K. For the fit in  Figure~\ref{fig:1}(d) we use the experimentally determined $\Theta_D$ of $1495$~K \cite{Pozzo:2011} and $\lambda = 8.8\times10^{-4}$.


Nevertheless, we can draw several important conclusions. The first is that $\lambda$ is very small, between $4\times10^{-4}$ and $2\times10^{-3}$. To the best of our knowledge, this is the lowest $\lambda$ value ever determined for any material. The result is consistent with the  the theoretical expectation of a vanishing $\lambda$ near the Dirac point \cite{Calandra:2007}, and with a single recent ARPES study for weakly doped graphene on SiC  \cite{Forti:2011}. Most earlier ARPES studies have been carried out for significantly stronger doping and have accordingly found   higher $\lambda$ values \cite{Bostwick:2007,Bostwick:2007b,Bianchi:2010}. The second conclusion is that the actual Debye temperature of graphene, rather than the Bloch-Gr\"uneisen temperature, appears to be the relevant temperature for the electron-phonon scattering. Again, the uncertainty of $\Theta_D$ in the fit is large because of the correlation between $\Theta_D$ and $\lambda$ but the fit is significantly inferior  for $\Theta_D$ values below 1050~K. $\Theta_{BG}$, on the other hand,  can be estimated to be $\approx 400$~K, using the average binding energy of 400~meV below $E_D$ that was used for the extraction of the temperature-dependent data and following Ref. \cite{Efetov:2010}.

While the electron-phonon coupling is thus consistent with theoretical expectations, the temperature-dependent changes of the electronic structure are highly unexpected. The most dramatic effect is the change from hole doping at low temperature to electron doping at high temperature. Indeed, if we infer the position of the Dirac point from an extrapolation of the occupied bands, its position changes by more than 250~meV over the temperature range explored here (see Figure \ref{fig:1}(f)).

It is tempting to ascribe this behaviour to an increased graphene-substrate interaction at higher temperatures. We have investigated this possibility by temperature-dependent \emph{ab initio} molecular dynamics calculations. In these calculations, a layer of graphene is placed on a three layer thick slab of Ir(111) and the atoms of the graphene and two topmost Ir layers are allowed to move for 60~ps, keeping track of the electronic degrees of freedom. Such calculations provide us with the average distance between the carbon atoms and the Ir(111) surface atoms and with the electronic structure of the entire system.

\begin{figure}
\begin{center}
\includegraphics[width=0.9\columnwidth]{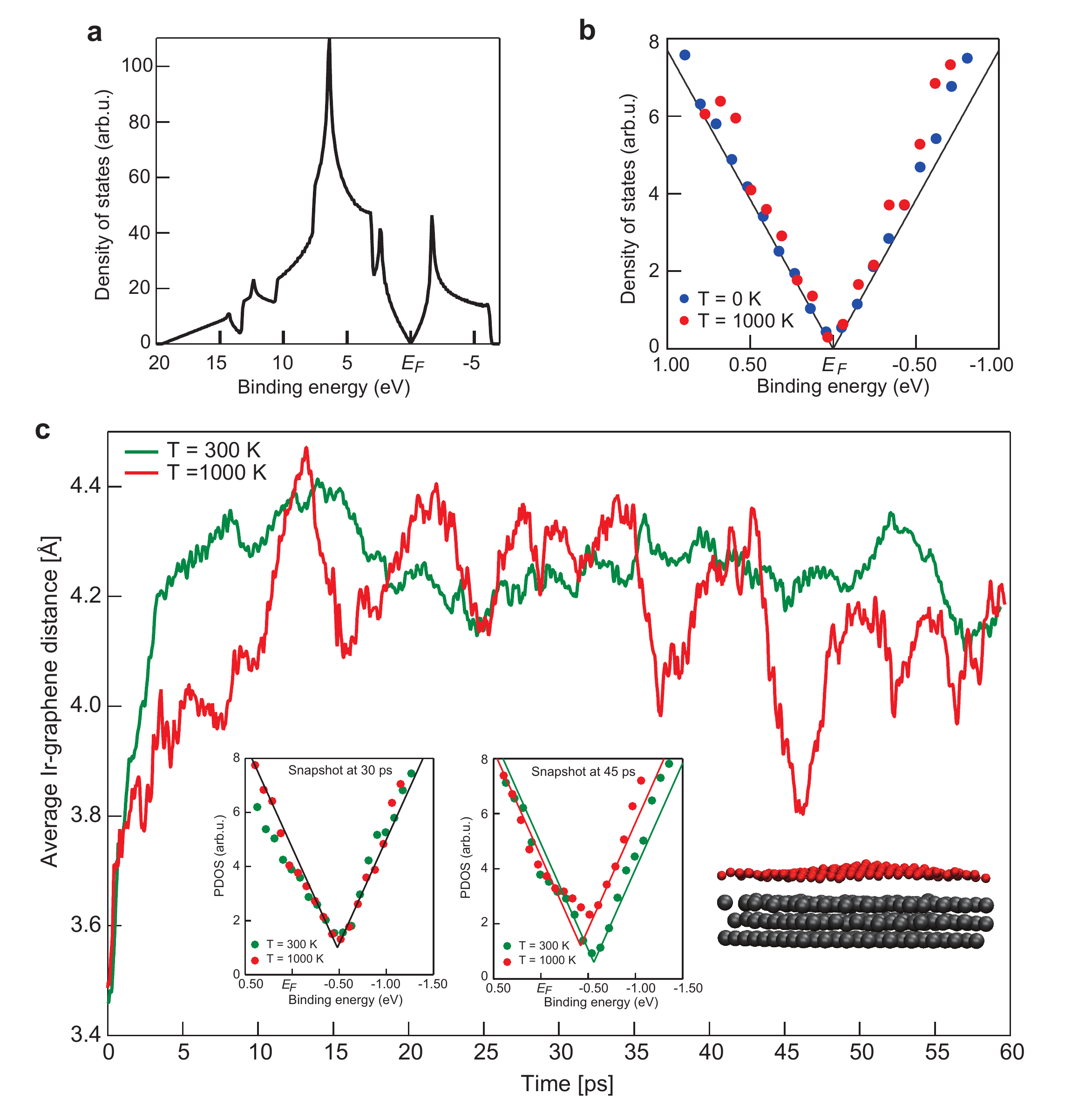}
\caption{Electronic and geometric structure of suspended and supported graphene determined by \emph{ab initio} molecular dynamics. (a) Density of states (DOS) of freely suspended graphene. (b) DOS for freely suspended graphene in the vicinity of the Fermi energy for two temperatures. The solid line is the DOS calculated from the analytic, linear dispersion. (c) Average distance between the Ir(111) surface and graphene during 60~ps simulations at 300~K and 1000~K. Insets: Snapshots of the projected DOS (PDOS) on the carbon atoms at 30~ps and 45~ps, corresponding to configurations with similar and difference graphene-Ir distances for the two temperatures, respectively. The solid lines have the same shape as in (b) but merely serve as a guide to the eye here. The geometry of the system after 45~ps at~1000 K is shown in the lower right corner.
\label{fig:2}}
\end{center}
\end{figure}

Figure \ref{fig:2}(a) gives the calculated density of states (DOS) for a freely suspended graphene layer at 0~K. It shows the expected features of a zero-gap semiconductor with the Dirac point energy $E_D$ at the Fermi energy. The electronic structure near $E_F$ is magnified in Figure \ref{fig:2}(b) and plotted together with the expected analytical result for a linear dispersion (solid line). The calculated and analytical results virtually coincide near $E_F$. Small deviations are only discernible for higher absolute binding energies, as the band structure becomes non-linear and the van Hove singularities, visible in Figure \ref{fig:2}(a), are approached. Also shown is the calculated DOS at a temperature of 1000~K. Remarkably, the temperature of the graphene has virtually no effect on the DOS.

The situation is dramatically different for supported graphene on Ir. Figure \ref{fig:2}(c) shows the result of a 60~ps \textit{ab initio} molecular dynamics calculation, giving the average graphene-Ir distance at 300~K and 1000~K. After some time needed to achieve thermal equilibrium (around 5~ps), the average distance fluctuates around a stable value. The distance fluctuations are much more pronounced at $T=$ 1000~K than at $T =$ 300~K, with the graphene layer coming closer to the Ir substrate. When this happens, the interaction between graphene and the substrate is stronger, resulting in a pronounced shift of $E_D$ towards higher binding energies. This is evident in the insets of Figure \ref{fig:2}(c) which show the projected density of states (PDOS) on the carbon atoms for two representative configurations: After 30~ps, the graphene-substrate distance is $\approx 4.2$~\AA~for both temperatures and the resulting PDOS curves are nearly identical. After 45~ps, however, the  average distance between the graphene layer and the substrate at 1000~K is reduced from $\approx 4.2$~\AA~ to $\approx 3.8$~\AA~and this results in a strong shift of $E_D$  from 560~meV to 420~meV above $E_F$. What is more, the PDOS is no longer well-described by the analytical model, contrary to suspended graphene, with the PDOS at $E_D$ now being substantially different from zero. 

The calculations thus reproduce and explain the ARPES observations: both the pronounced change in doping and the deviation of the spectral function from a simple Dirac cone are caused  by fluctuations in the graphene-substrate distance as the temperature is increased.  It is tempting to make a more quantitative comparison between experiment and calculations but such a comparison would exceed the achievable accuracy in the calculations. The main reason is that the average distance between graphene and the Ir substrate is probably overestimated by the calculations, which do not include the van der Waals interaction. This is consistent with the difference in doping between ARPES and calculation. The experimental trends are, however well reproduced and we believe the essential physics to be captured. Note that a fluctuating doping of graphene at high temperature would lead to a systematic error in our determination of $\lambda$ because it represents an additional broadening mechanism. This, however, would merely cause the real $\lambda$ to be even lower than the value we report above.

The observed temperature-dependent changes of the electronic structure are expected to lead to a very complex behaviour in transport measurements, even for a simple metallic substrate without any polar phonon modes. In fact, the contribution of the electron-phonon coupling would be expected to be insignificant with respect to the other changes that would presumably give rise to a ``semiconducting'' behaviour caused by a strong decrease of the carrier density between 0~K and 700~K and a ``metallic'' behaviour above 700~K. Most transport measurements are admittedly limited to a much smaller temperature range but our results illustrate that that the temperature-dependent doping of supported graphene could have a very significant impact on the transport properties. 

In conclusion, we have used spectroscopic measurements showing that the electron-phonon coupling for supported graphene can be extremely weak. Nevertheless, strong effects in the temperature-dependent transport properties can be expected due to temperature-dependent doping changes of the graphene. Our results are specifically important for a graphene-metal interface where the doping of the graphene has important consequences for device operation \cite{Wu:2012}. But they are not restricted to this type of interface. Graphene on SiO$_2$ is also subject to considerable interface charge transfer \cite{Romero:2008} and similar effects can be expected. Finally, we note that pristine and suspended graphene could be expected to retain its benign electronic properties up to very high temperatures, as our results suggest that the intrinsic electron-phonon coupling is very weak indeed and thermal fluctuations would hardly affect the DOS. 

\section{Methods}

ARPES experiments were carried out at the SGM-3 beamline of the synchrotron radiation source ASTRID \cite{Hoffmann:2004}. Graphene was prepared on Ir(111) using a well
established procedure based on C$_2$H$_4$ dissociation \cite{Coraux:2009}. The quality of the graphene layer was controlled by low-energy electron diffraction and its spectral function was measured by ARPES. At low temperature, the photoemission linewidth of the features was found to be similar to published values \cite{Kralj:2011,Pletikosic:2009}. The temperature measurements were performed with a K-type thermocouple and an infrared pyrometer. The temperature-dependent data were taken such that the sample was heated by a filament mounted behind it. The filament current was pulsed and the data were acquired during the off-part of the heating cycle. The total energy and $k$ resolution during data acquisition were 18~meV and 0.01~\AA$^{-1}$, respectively. The MDC linewidth was determined as the average over an energy range between 250~meV and 550~meV below the Dirac point. An energy interval was chosen in order to improve the experimental uncertainties. The interval limits were chosen such that the lower limit is always more than a typical phonon energy ($\approx 200$~meV) away from $E_F$ and neither limit is too close to  the Dirac point or the crossing points between the main Dirac cone and the replica bands, as this is known to lead to errors in the linewidth determination \cite{Nechaev:2009}.

The {\it ab initio} calculations were performed with the VASP
code~\cite{Kresse:1996}, the projector-augmented-wave method~\cite{Blochl:1994,Kresse:1999}, the
Perdew-Burke-Ernzerhof exchange-correlation energy~\cite{Perdew:1996}, and an
efficient extrapolation for the charge density~\cite{Alfe:1999}. Single
particle orbitals were expanded in plane waves with a cutoff of 400
eV. We used the NPT ensemble (constant particles number $N$, pressure $P$,
and temperature $T$), as recently implemented in
VASP~\cite{Hernandez:2001,Hernandez:2010}. For the present slab
calculations, we only applied the constant pressure algorithm to the
two lattice vectors parallel to the surface, leaving the third
unchanged during the simulation. Adsorption of graphene has been
modelled by overlaying a $10\times 10$ graphene sheet (200 C atoms)
over a $9 \times 9$ Ir(111) supercell~\cite{Pozzo:2011} and using a slab
of 3 layers where the two topmost layers were allowed to move while
the bottom layer was kept fixed. Molecular dynamics simulations were
performed with the $\Gamma $ point only at T = 300 K and T= 1000
K. Density of states were calculated on representative simulation
snapshots, using a $16 \times 16 \times 1$ grid of {\bf k}-points (128
points). The projected density of states on the carbon atoms were obtained by projecting the Bloch orbitals onto spherical harmonics with $l=$1, inside spheres of radius 0.86~\AA~ centered on the C atoms. The PDOS obtained in this way is representative of the density of states due to the $p$ orbitals of the carbon atoms.

The DOS of suspended graphene at 0 K was calculated only for the $p$ states, as the $s$ state contribution around the Fermi energy is very small. The DOS was rescaled such that it could be fitted to the analytical linear density of states prer unit cell of isolated graphene near the Dirac point. The same scaling factor was applied to all the calculated density of states data. The position of the Dirac point in the calculations for supported graphene was determined by fitting the analytical linear DOS with an offset in binding energy corresponding to the new position of the Dirac point. 
\section{Acknowledgements}
This work was supported by The Danish Council for Independent Research / Technology and Production Sciences and the Lundbeck foundation.  A.B. acknowledges the Universit\`a
degli Studi di Trieste for the \textit{Finanziamento per Ricercatori di Ateneo}-FRA2009.  The \textit{ab-initio} calculations were performed on the HECToR national service in the U.K.


\end{document}